# Synthesis and characterisation of superconducting $RuSr_2GdCu_2O_8$

**C.Artini**[1], **M.M.Carnasciali**[1], **G.A.Costa**[1], **M.Ferretti**[1], **M.R.Cimberle**[2], **M.Putti**[3], **R.Masini**[4,*]

*1 INFM and DCCI, Università di Genova, Via Dodecaneso 31 – 16146 Genova (Italy)*
*2 CNR CFSBT, Via Dodecaneso 33 – 16146 Genova (Italy)*
*3 INFM and DIFI, Università di Genova, Via Dodecaneso 33 – 16146 Genova (Italy)*
*4 CNR TeMPE, Via Cozzi 53 – 20125 Milano (Italy)*

\* corresponding author. Tel.: +39 02 66173 337; Fax +39 02 66173 321. E-mail address: masini@tempe.mi.cnr.it

**Abstract:**

We report on the structural, electrical and magnetic properties of $RuSr_2GdCu_2O_8$ samples made by systematic synthesis work so that they differ from one another in number of annealings and sintering temperature. Our aim is to highlight the different behaviours as a consequence of the thermal treatments. In particular, we deal with the onset of superconductivity in relation to the homogenisation process, and its influence on resistivity, structural and magnetic ordering.

Finally, the problems related to the magnetic measurements due to the superimposition of different magnetic states are examined.

**Keywords:** Ru1212, synthesis, superconductivity, electrical and magnetic properties.

## 1. Introduction

$RuSr_2GdCu_2O_8$ ruthenocuprate (Ru-1212) belongs to a class of compounds that were synthesized for the first time in 1995 [1]. It is characterised by a triple perovskitic cell similar to that of $YBa_2Cu_3O_x$, where Y and Ba are substituted by Gd and Sr, respectively, and the $CuO_2$ planes are separated from one another not by the Cu chains, but by $RuO_6$ octahedra.

This first synthesis work yielded a material whose superconductivity, verified only through electrical measurements, could not be undoubtedly attributed to the 1212 phase. Since the discovery of the coexistence of ferromagnetism and superconductivity in this class of materials, the efforts of many researchers have been focused on this phase. Nevertheless Chu et al. [2] still question about the actual existence of bulk superconductivity.

Since it is well known that physical properties are strongly dependent on sample preparation conditions, the critical point of all investigations on these materials is the preparation method. The need to avoid the formation of impurities, such as the ferromagnetic $SrRuO_3$ for example, and to obtain a phase as ordered as possible, led to the optimisation of a synthesis process that includes a first calcination, a thermal treatment in $N_2$ [3,4] and several annealings in flowing $O_2$ [4,5,6].

Reports of both superconducting [4,6] and non superconducting samples [7,8] of Ru-1212 have spurred a detailed experimental approach aimed to determine the factors that influence superconductivity of this material. Tackling the question of how the superconducting and magnetic properties are affected by subtle details of the preparation process, which in turn affect the microscopic structure, we present the results of a systematic work on the synthesis of Ru-1212 which allows superconductivity to be induced in a controlled manner.

## 2. Experimental details

Polycrystalline samples of composition $RuSr_2GdCu_2O_8$ have been synthesized by solid state reaction of high purity stoichiometric powders of $RuO_2$, $Gd_2O_3$, CuO and $SrCO_3$. The mixture was first calcined in air at 950°C; then, after grinding and die-pressing into pellets, it was treated in flowing $N_2$ atmosphere at 1010°C. This step resulted in the formation of a mixture of the precursor materials $Sr_2GdRuO_6$ and $Cu_2O$ without formation of the impurity phase $SrRuO_3$, very stable in oxidising environment [3]. The mixture was then subjected to eight successive sintering steps in flowing $O_2$, each one lasting 15 hours, at temperatures starting from 1030°C up to 1085°C. Each successive thermal treatment was performed at a temperature 7°C higher than the previous one. At the end of every step the product was quenched to room temperature, fully characterized, reground, pressed into pellets and subjected to the successive treatment at higher temperature, so that the effects of the thermal treatments during synthesis process could be investigated. Each reaction step was carried out on a MgO single crystal substrate to prevent reaction with the alumina crucible. This procedure allowed us to highlight the passage from a non superconducting to a superconducting status in the material.

In this work, we report on the structural, electrical and magnetic properties of three samples considered indicative of the overall process of synthesis and formation of the Ru-1212 phase. Particularly those which were subjected to:

(i) three sintering steps up to 1045°C as follows: 1030°C x 15h + 1037°C x 15h + 1045°C x 15h, for a total time t= (15 x 3) = 45h,

(ii) five sintering steps up to 1060°C for a total t = (15 x 5) = 75 h,

(iii) eight sintering steps up to 1085°C for a total t = (15 x 8) = 120 h;

hereafter referred to as samples A, B and C, respectively.

The crystal structure of the samples was examined by X-ray powder diffraction using CuK$\alpha$ radiation. Micro-Raman analysis was carried out by a Renishaw System 2000 instrument (He-Ne laser, 633 nm, 15mW, 50x, 4 accumulations). The resistivity of sintered polycrystalline bars was measured by the standard four-probe technique with a 1mA current in the 15 – 300 K temperature range. Magnetic susceptibility was measured in a Quantum Design SQUID magnetometer.

**3. Results and discussion**

3.1. *Structure and microstructure*

XRD patterns showed single phase 1212-type materials with no traces of impurities within the resolution of the technique. All diffraction peaks could be indexed on the basis of a tetragonal lattice (Table 1) in agreement with the literature data [3]. In order to test the phase homogeneity furtherly and due to its greater sensitivity with respect to XRD technique, micro-Raman analysis has been performed. The typical spectra obtained in three different points of sample B are reported in Fig. 1, where modest but detectable differences can be observed. The overall behaviour is in agreement with the literature data [9] showing the same structures of the peaks: the slight change of the barycentre and band amplitude are presumably due to the different crystallinity between the centre and the border of the sample. A broadening and a small shift toward lower frequency of the peak at 640 cm$^{-1}$ (apical O(Sr) $A_{1g}$ mode) have been detected in A and C sample spectra which can be related to a softening of the Sr-O bond. A comprehensive discussion and analysis of micro-Raman spectra in these compounds will be the subject of a future work.

SEM analyses show high grain homogeneity with clean grain boundaries. Moreover, we have detected a progressive grain growth with a corresponding increase in grain connectivity due to the different thermal treatments. Complete sintering was never reached. This fact can be explained considering the difference between the decomposition temperature of the 1212 phase ( around 1130°C, as estimated through thermal analysis) and the temperature of the thermal process ($T_{max}$ = 1085°C in this work). The geometric density varies by about 5% between the first and the final bulk sample C, which has a density of 4.2 g/cm$^3$ (about 63% of the theoretical crystallographic density).

In order to investigate whether the observed behaviours were caused by the annealing time or temperature, tests have been performed by comparing the properties of the samples obtained through the successive sintering steps mentioned above with other samples, of the same composition and batch, subjected to a single thermal treatment at the final temperature of each single step for a time which is the sum of the corresponding partial times of the serie considered in this work. This allowed us to conclude that reaching the "optimal" temperature directly does not produce the same results of the longer procedure: repeated homogenisations, related to the sequence of grinding and annealing, controls the superconducting behaviour.

*3.2. Electrical properties*

The behaviour of the room temperature resistivity as a function of the number of sintering steps n is reported in Fig. 2. It can be clearly seen that $\rho_{300}$ has an almost constant value at the beginning of the series, then it suddenly drops, reaching another constant value that is about six times smaller than the starting one. Such different values are not related to the sample density, which does not present strong variations. In the samples with n ≤ 4, a semiconductor-type behaviour (with a local minimum at 130K) is observed in the 50-300K temperature range. At higher n, a metallic type behaviour is detected, with a well defined minimum at 130K followed by a semiconducting upturn. The minimum at T=130 K is related to the onset of the magnetic ordering of Ru lattice. Moreover, for n<4 a maximum in resistivity is observed at 35K, which is progressively depressed in magnitude and shifted up to 45K at higher n. At lower temperatures a large resistivity drop (down to values 100 times smaller than $\rho_{300}$) is measured for all samples with n<4. At higher n superconductivity sets in and zero resistivity is detected. In view of this trend in resistivity we limit our treatment to sample A (n=3), whose behaviour lies in the non superconducting and semiconductive region, to sample B (n=5), for which superconductivity and metallic character in resistivity have just set in, and to sample C (n=8), for which superconductivity has been depressed. Resistivity measurements performed on these three samples are reported in Fig. 3. The temperature derivative of the resistivity for sample B is shown in the inset of Fig.3. A two-step behaviour is observed: a high temperature peak at about 45K, corresponding to the superconducting thermodynamic critical temperature, and another one at lower temperature (here at about 30K). This behaviour is commonly observed in the literature: different explanations have been proposed [10,11].

3.3. *Magnetic properties*

An experimental problem encountered in the magnetic measurements of Ru-1212 materials is strictly related to the complexity of the magnetic signal. As noted in [12], the magnetization in Ru-1212 contains magnetic moments arising from different contributions: the Gd paramagnetic spin lattice, the Ru spin lattice and, finally, the diamagnetic signal related to the superconducting behaviour. For both the Gd and Ru spin lattice the antiferromagnetic ordering is coupled with a ferromagnetic component that is attributed to a canting of the lattice, in the case of Ru, whereas, in the case of Gd, is simply related to the presence of the net ferromagnetic moment of the Ru lattice [13]. Therefore, the basic condition of a homogeneous magnetic moment requested by the SQUID magnetometer is not fulfilled, in particular at low temperatures where magnetic moments of opposite polarity will be present in the sample as a consequence of the applied field. In addition, in the superconducting coil of the magnetometer a small remanent field may be present. This last can be zeroed with a proper procedure which results in a residual field of the order of fraction of Gauss (estimated by measuring the susceptibility of a paramagnetic salt) in the central part of the magnet, and greater far from it. Therefore, a "real" ZFC measurement cannot be made and, since the FC magnetic moment is about one order of magnitude greater than the ZFC one, even a residual field of a fraction of Gauss may give rise to a considerable magnetic signal whose polarity depends on the field polarity. Finally, we point out that, when the sample is moved for a length that is usually of few centimetres during the measurement, it moves in a non uniform magnetic field that makes it follow a minor hysteresis loop. If the value of the moment is not constant during the scan, an asymmetric scan wave form will be detected, and the quality of the measurement will drastically degrade [14].

The ZFC and FC measurements are reported in Fig. 4 a) and b). The magnetic ordering manifests itself as a cusp related to an antiferromagnetic ordering in the ZFC measurement, whereas, in the FC curve, as a sudden onset of a spontaneous magnetic moment related to a ferromagnetic component. For all three samples the spontaneous magnetization occurs in the 130-135K temperature range. At the lowest temperatures a contribution from Gd sublattice, which orders antiferromagnetically at 2.5K, is clearly visible. The superconducting behaviour of sample B is detected by the diamagnetic shift corresponding to the shielding of about 50% of sample volume at the lowest temperatures and fields. Measurements performed on sample B at different applied fields are shown in Fig. 4b). The FC susceptibility as well as the amplitude of the irreversibility between FC and ZFC measurements decreases as the field increases. At the minimum applied field of 0.2 G, and to a minor extent even at 1.2 G, a Meissner behaviour is seen in the FC curve that quickly re-enters. At 5.5 G the Meissner effect is only seen as a constant value hindering the Gd magnetic ordering. The poor visibility of the Meissner effect has been observed by several authors [10,15] and explained in different ways: Bernhard et al. [10] have proposed that a spontaneous magnetisation from $RuO_2$ layers results in a local magnetic field that may be greater than the lower critical field $H_{c1}$ giving rise to a spontaneous vortex phase (SVP). The effect of a SVP is to suppress the Meissner phase. Chu et al [15] have associated this behaviour to a phase-lock of an aggregation of small Josephson-coupled superconducting grains or domains.

In Fig. 5 the high temperature susceptibility data at 1000 G are reported. The best fit to the experimental data with a two components curve $\chi=C_{Gd}/T +C_{Ru}/(T-\theta)$ is shown and the fitting results are reported in Table 1. Shown in the inset is the M(H) curve measured at 5K. With our maximum available field, saturation of the magnetisation is not achieved.

**Conclusions**

It has been demonstrated that bulk superconductivity can also be obtained in quenched $RuSr_2GdCu_2O_8$ samples. Superconductivity, as probed by magnetic susceptibility, is observed after proper homogenisation. At the higher temperatures (T>1070°C) the annealing tends to introduce disorder in the system, worsening the superconducting properties.

In properly homogenised samples we have been able to observe a 50% sample volume shielding and a magnetic signal related to a Meissner effect.

## Acknowledgements

This work was partially supported by INFM and CNR.

**Figure captions**

**Fig. 1** Micro Raman spectra obtained in three different points of sample B.

**Fig. 2** Resistivity at 300 K as a function of the number of sintering steps. Hatched region separates zones which show progressively different electrical behaviours.

**Fig. 3** Resistivity of samples A ( ), B ( $\Delta$ ) and C ( O ) as a function of the temperature. Inset: temperature derivative of resistivity vs. T for sample B in the transition region.

**Fig. 4** ZFC (hollow symbols) and FC (filled symbols) magnetic susceptibility as a function of temperature (a) for samples A ( $\Diamond$ ) at H = 3.5G and C ( $\Delta$ ) at H = 1.5G; (b) for sample B.

**Fig. 5** Magnetic susceptibility vs. temperature for samples A ( ), B ( $\Delta$ ) and C ( O ) in the paramagnetic region. Fitting results are also shown as solid lines. Inset: M(H) curve at 5K for sample B.

**Table 1** Synthesis, structural, electrical and magnetic data of our $RuSr_2GdCu_2O_8$ samples. $C_{Gd}$, $C_{Ru}$ and $\theta$ come out from a best fit of the susceptibility experimental data in the 200-300K temperature range with a two component function.
.

**Table 1**

| Sample | $T_{sintering}$ (°C) | a (Å) | c (Å) | $\rho_{300}$ (mΩ cm) | $T_{max}^{ZFC}$ (K) | $C_{Gd}$ (emu K/mole) | $C_{Ru}$ (emu K/mole) | θ (K) |
|---|---|---|---|---|---|---|---|---|
| A | 1045 | 3.831(1) | 11.545(7) | 137.4 | 130 | 7.38 | 1.70 | 108.6 |
| B | 1060 | 3.828(1) | 11.552(6) | 30.7 | 130 | 7.60 | 0.79 | 144.9 |
| C | 1085 | 3.835(1) | 11.580(5) | 22.0 | 135 | 8.26 | 0.81 | 139.6 |

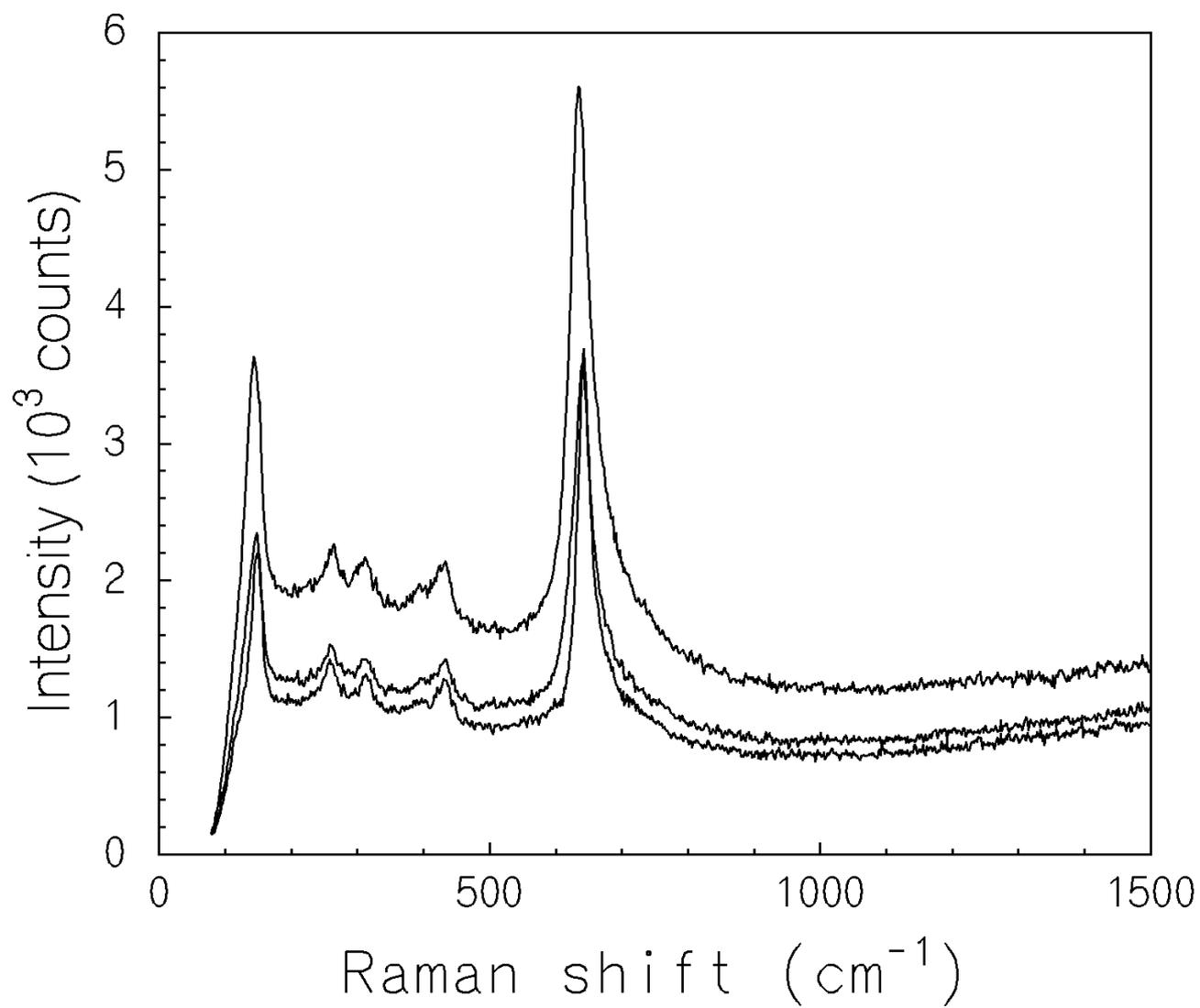

Fig. 1

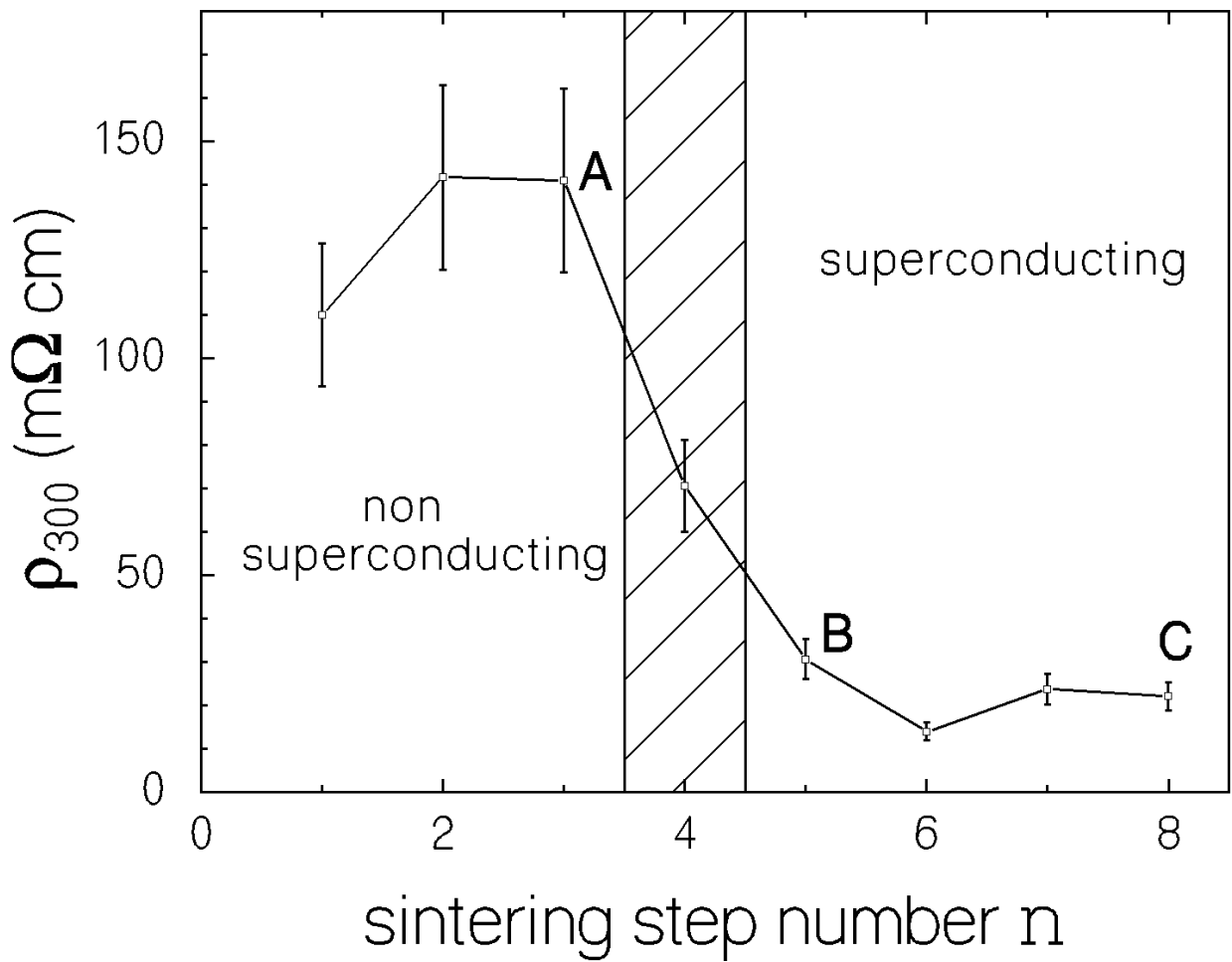

Fig. 2

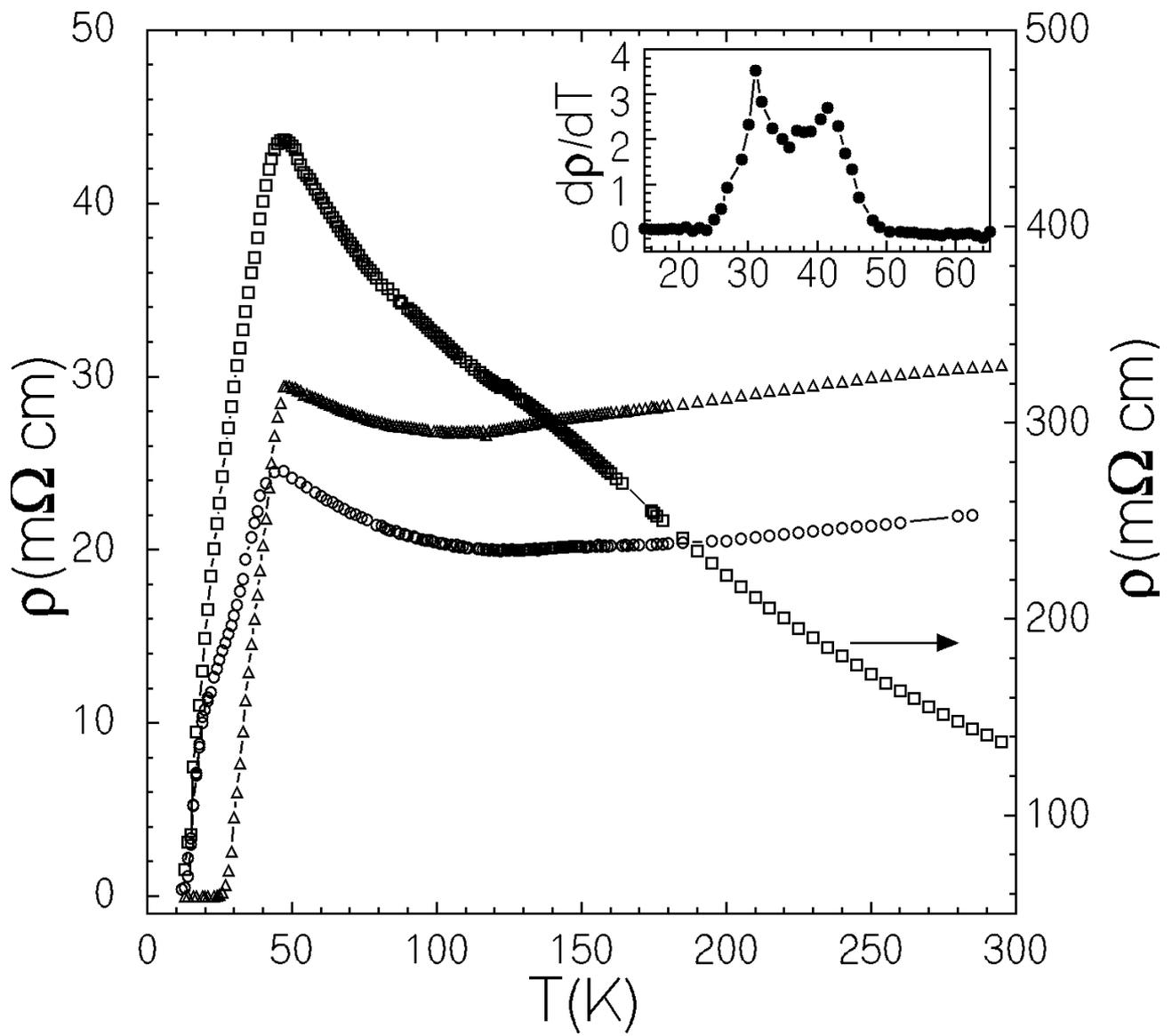

Fig. 3

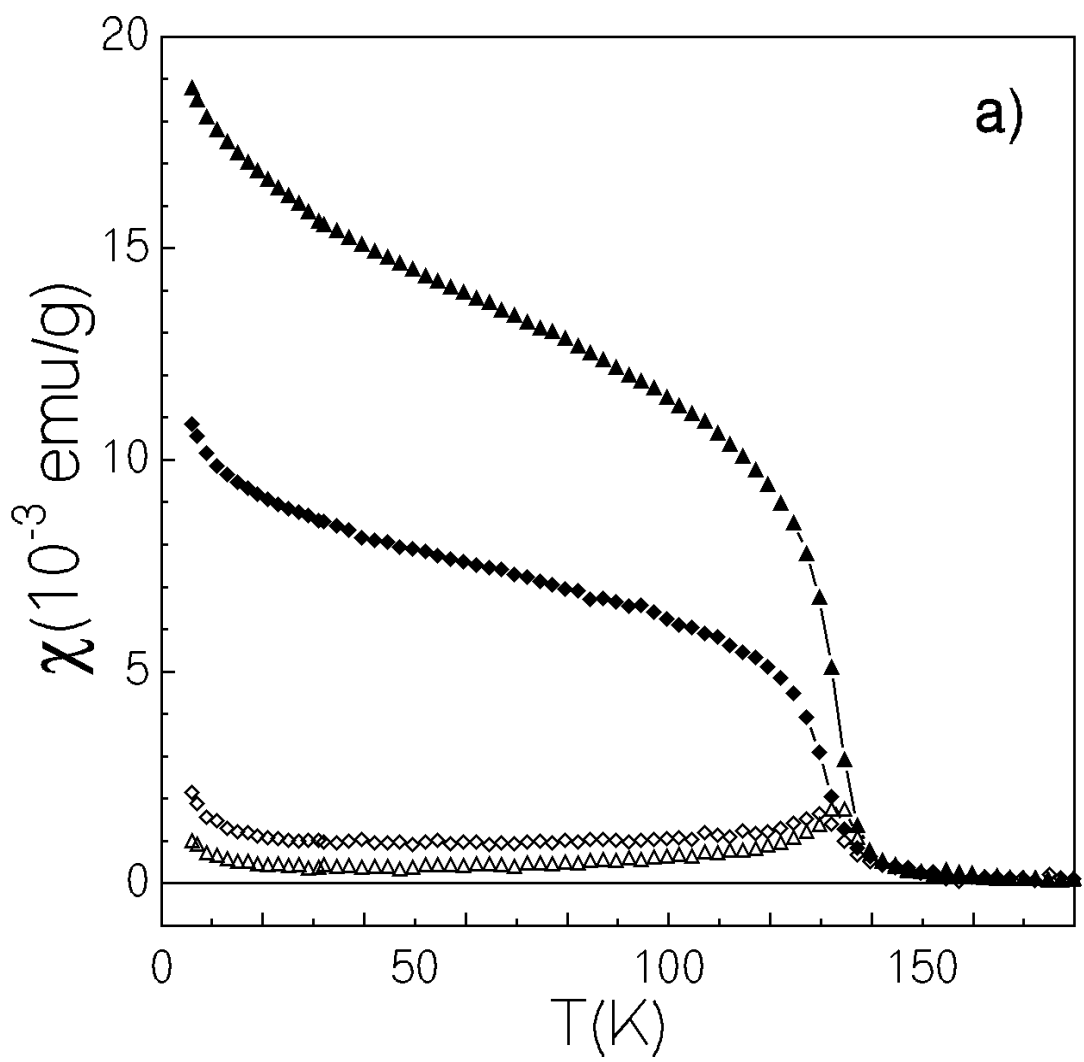

Fig. 4a

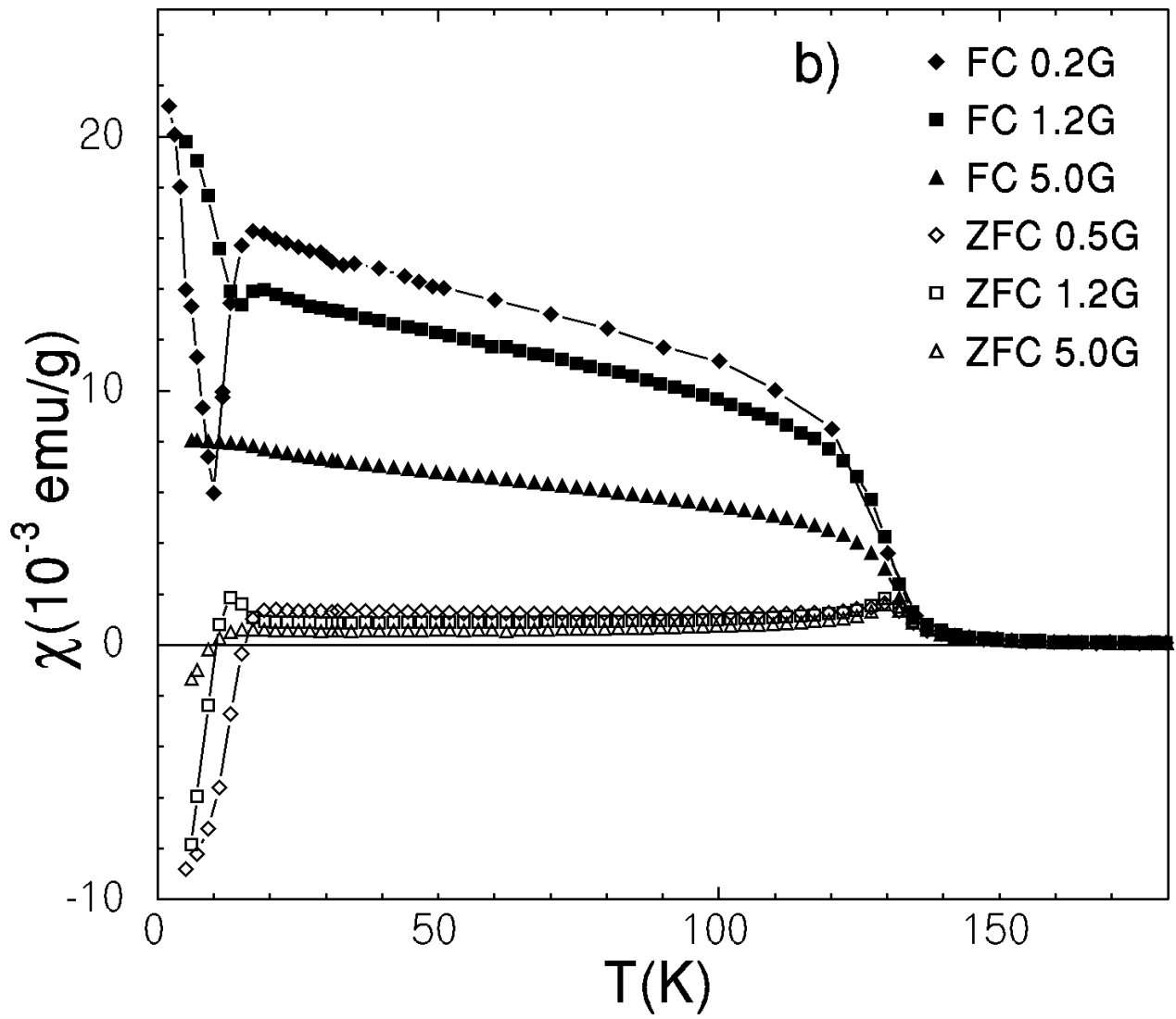

Fig. 4b

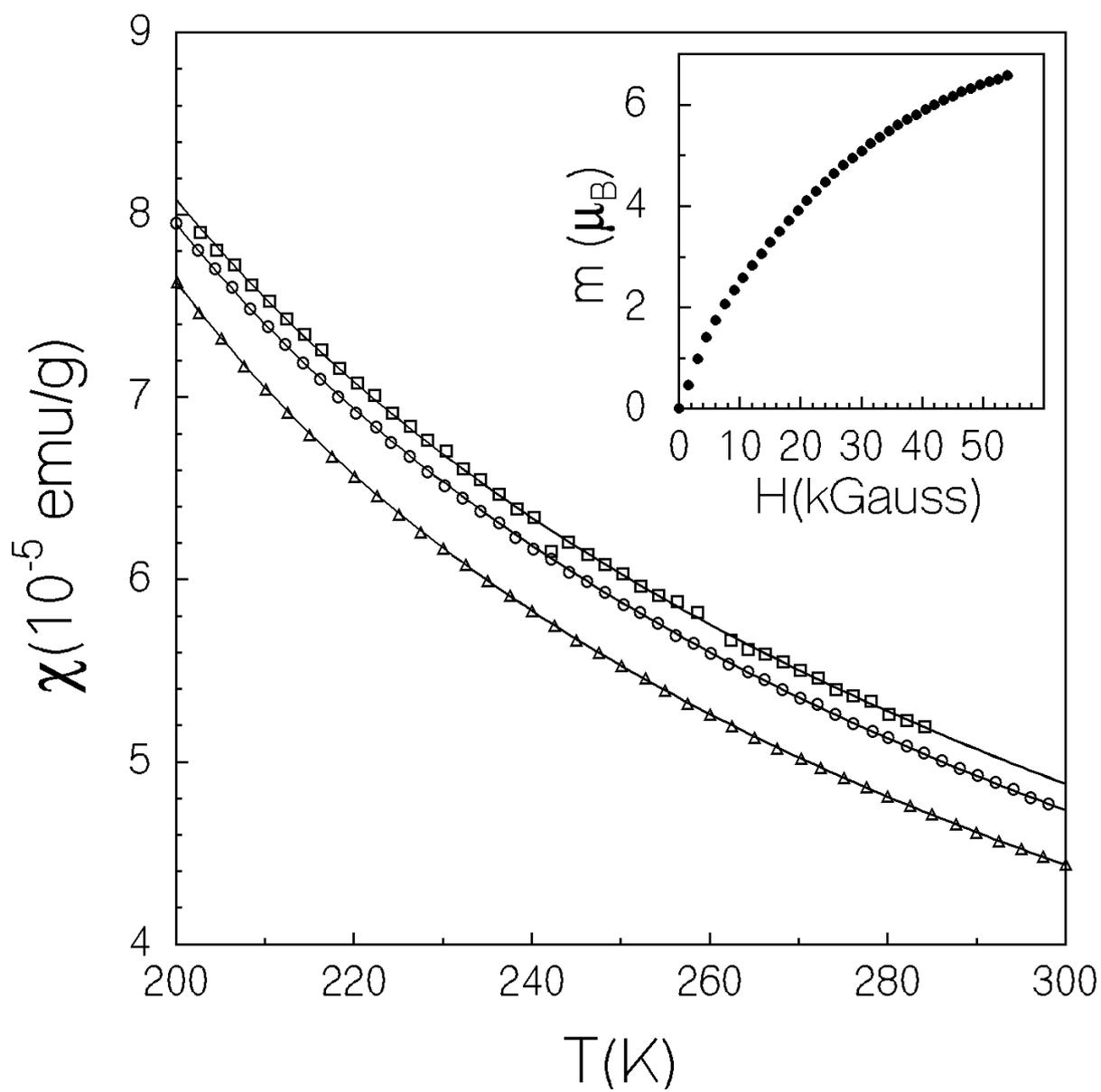

Fig. 5